\documentstyle[12pt]{article}
\newcommand{\pp} {\vskip 0.08in}
\renewcommand{\baselinestretch}{1.9}
\begin{document}
\renewcommand{\baselinestretch}{1.9}

%\title{
\begin{center}
\begin{Large}
Water Droplet Avalanches
\end{Large}
\end{center}

%\maketitle
\vskip 0.3in
\begin{center}
Britton Plourde, Franco Nori and Michael Bretz \\
{\it Department of Physics, The University of Michigan, Ann Arbor, MI
48109-1120}
\end{center}

\vskip 0.3in
\begin{abstract}
We analyze the statistics of water droplet avalanches in a continuously
driven system.  Distributions are obtained for avalanche size, lifetime,
and time between successive avalanches, along with power spectra and
return maps.
For low flow rates and different water viscosities, we observe a
power-law scaling in the size and lifetime distributions of water droplet
avalanches, indicating that a state with no characteristic time and length
scales was reached.
Higher flow rates resulted in an exponential behavior with characteristic
scales.

\end{abstract}

\pp
\vskip 0.3in
Accepted in Phys. Rev. Lett.
%\section{Introduction.}

\newpage
%\twocolumn

The deposition, growth, coalescence and motion of fluid droplets
is a subject of enormous interest to many disciplines in science and
engineering\cite{reviews}.  Examples of fluid transport that have recently
attracted renewed interest include models of river formation\cite{rivers} and
the coverage area fluctuations of periodically-pulsed droplet
deposition\cite{rain}.
\ Also, studies of a {\it single\/} dripping faucet have found a
rich dynamics, including period-doubling, as a function of the water
supply flow rate\cite{faucets}.  Lightning discharges and other forms of
electrical breakdown\cite{weiss}, aperiodic X-ray sources\cite{crutchfield},
plasma turbulence, mechanical vibrations, and electronic circuits\cite{nrl}
can be sometimes characterized in terms of
spatially-extended ``dripping" models.  The literature on water droplet
dynamics is vast, and we do not attempt a review.  The interested reader
is referred to our references and papers cited therein.   Here, we
present an investigation of the dynamics of avalanches in
continuously driven water droplet systems.

\pp
Recently, the notion of self-organized criticality has been
proposed\cite{btw}
to account for the behavior of certain driven dissipative dynamical systems
which exhibit long-range spatial and temporal correlations.  Upon reaching
the critical attractor,
avalanches are produced with lifetime and size distributions which decay as
power laws.  Many
previous avalanche studies have examined the onset of collective transport in
a wide variety of systems in which non-coalescing interacting elements
(e.g., sand grains, vortices) can be identified.  Examples include
granular materials\cite{sand} and vortices in superconductors\cite{vortices}.
Our study is one of the few systematic investigations of collective dynamics
in a system with coalescing moving elements (i.e., a fluid).
%, in contrast with examples with
%non-coalescing moving elements (e.g., sand grains, vortices).
\ An avalanche on a tilted sprayed surface occurs when a droplet grows in size
and eventually reaches a critical mass, at which time it runs down the surface,
triggering other stationary droplets in its path, thus creating a chain
reaction\cite{meakin}.
\ Here, we examine the power spectra and distributions of water droplet
avalanche sizes, lifetimes, and delay times between successive events
in an analogous manner to previous studies of avalanches.

%{\it Experimental Apparatus\/}

\pp
The experimental apparatus is illustrated in fig.~1(a).  Distilled water
was sprayed through a flow-regulated forced-air mister into a clean,
clear plastic dome.
\ The water spray was collected inside the dome, producing streams
which ran down and eventually dripped off the rim.  The droplets then
impacted on, and immediately ran off of a light slanted ring,
%made of circuit board (G-10) material,
which was suspended by a piezo film detection system.
%, finally falling to a pan.

\pp
Figure 1(b) displays the anatomy of one of the three piezo film detectors.
A drop of water falling on the
ring stretched the film longitudinally, thus producing a measurable voltage
from which we determined the time stamp and size of the
water droplet avalanches.  Each piezo strip transducer consisted of a thin
Mylar film of double-coated piezoelectric material, which was
hung within a copper housing.
\ Each of the three signals was sharpened by an RC filter, passed through an
amplifier circuit,
and was digitized using a Data Translation 2805 converter board.  Data were
acquired at a sampling rate of $4$ kHz using a double-buffered
direct memory access A/D conversion technique which allowed for a
continuous flow of data at $1.33$ kHz % ($=4/3 \; kHz$)
per channel.
% This involved using direct memory access control to input
%\ Data went to the first half of the buffer array, while the computer
%simultaneously processed data in the other buffer half.

\pp
Each droplet impacting the ring produced direct signal responses followed by
post-impact (or aftershock) damping vibration signals in all three detectors.
% along with post-impact damping vibrations (aftershocks).
\ A program was written in $C^{++}$ to compare the detector signals with
pipette calibrated impulses, to test a cut-off value for screening
out aftershocks, to sum the signals, and to output a time stamp and
relative size of each event.  Calculation of the relative event size
was possible because the total detector signal produced
by one fixed size droplet was essentially independent of angular position
around the ring.
%  (for longer version) This is illustrated in fig 1(c) where the total
% signal strength is plotted against the angular displacement for $0.2$ ml
% droplets from a pipette.
%\ A pipette was used to calibrate the average droplet signal at each
%detector.  Cutoff values which were too large caused the program to miss
%smaller events, while values which were too small caused the program
%to include aftershocks as actual events.
%\ Due to the impulse-sensing
%nature of this detection method, the avalanche size actually corresponds
%to a flow of droplet momenta, rather than to a mass flow only.

\pp
Because this detection system could discern droplets separated in time by only
a few milliseconds, it was necessary to define a clustering time,
$T_{cl}$, for droplets that were considered to be causally connected to each
other, and not independently triggered.  The individual droplet signals
(with a duration on the order of $0.5$ msec) separated by less than $T_{cl}$
were considered to be part of the same avalanche (i.e., an event composed of
one or more individual droplets), with a
total size equal to the sum of the sizes of the constituent
individual droplets.  The time stamp was assigned as that of the
initial droplet time.  The value of $T_{cl}$ could be easily altered
in the analysis program, allowing us to test many different values on the
same raw data set and to view the effects.
\ For $T_{cl} \leq 5 $ msec, droplets which were observed to be correlated,
were artificially subdivided into separate events.
\ For $T_{cl} > 7 $ msec, a visual
observation of the output indicated that the evaluation program summed many
small uncorrelated events, lumping them together as larger avalanches.
\ Therefore, we have chosen $T_{cl}= 6 $ msec.
\ This evaluation
method also made it straightforward to define an avalanche lifetime, $\tau$,
as the time from the initial to the final droplet in a given avalanche, and
also a delay time, $\Delta t$, between avalanches.

\pp
The system produced many water streams which ran down the surface, often
resulting in water accumulations around the rim of the dome.  %From them,
Drops then fell off the rim and impacted the ring when they acquired the
critical mass needed to overcome surface tension effects of the edge,
or when a fast moving droplet hit an accumulation and triggered an avalanche.
\ Since the detection method could only monitor avalanches that actually
fell off the dome, it was decided to minimize this accumulation problem
by attaching many $1$ cm long vertical copper wires along the rim, thus
preventing rim droplets from coalescing and lingering. % at the rim.
\ In early stages of this experiment, mechanical vibrations were a significant
problem.
\ The dome base was therefore fit with a thick brass outer sleeve and tightly
wedged in a mechanically rigid frame, thereby removing vibrations.

\pp
In order to examine the effects of increasing the water viscosity, the water
temperature was lowered. By chilling water from room temperature to near
freezing, the water viscosity approximately doubles.  At $25^{\circ}$C
the viscosity of water is 0.89 cp, while at $1^{\circ}$C the viscosity is
1.73 cp.   This method of varying the viscosity appeared to be simpler
and more controlled than using a fluid other than water.  Cooling was
accomplished by adding chilling elements to the dome, spray nozzle,
and water reservoir.

%\pp
%{\it Results}

Data runs were taken at many different flow rates, for both high and low
temperatures.
\ The results shown here represent two data runs of 3.5 hours each
with a low flow rate (8 cc/min):
one at about $22^{\circ}$C and one near $1^{\circ}$C (hereafter denoted as
``high T'' and ``low T'', respectively).
Figure 2 contains the distributions of avalanche sizes, $D(S)$, and
lifetimes, $D(\tau)$.  Figure 3 shows the corresponding distributions of
time delays between consecutive avalanches (sometimes called waiting times),
$D(\Delta t)$.
\ The plots have been scaled so that the sum
of the probabilities for each data set is equal to one.  All of our data
were examined on semi-log and
log-log plots, with both linearly and logarithmically binned histograms.
\ The latter have bin heights which were normalized by their corresponding
bin widths.
\ These alternate plotting and binning methods helped us to discern
between exponential, power-law, or other types of decays.

\pp
The low flow rate size distributions, fig.~2(a), exhibit a power-law behavior,
as shown by the linear fits:
$D(S) \sim S^{-a}$, with $a=2.66$ ($1.93$) for low (high) viscosity.
\ The high flow rate size distributions (not shown) exhibit an exponential
behavior: $D(S) \sim \exp \{ -(S/S_0) \}$, with $S_0 \sim 1.2$, consistent
with single droplets.
\ $D(\tau)$ for the high T data, fig.~2(b), extended over a somewhat narrow
range of $\tau$ due to the low water viscosity.  By lowering the temperature,
and thereby
increasing the ``cohesive strength'' between the droplets (i.e., water
viscosity), $D(\tau)$ extends over a longer $\tau$ range.
\ The linear fit in fig.~2(b) indicates a power-law behavior over about a
decade: $D(\tau) \sim \tau^{-3.10}$.
\ The high flow rate lifetime distributions (not shown) exhibit an exponential
behavior: $D(\tau) \sim \exp \{ -(\tau/\tau_0) \}$, with $\tau_0 \sim 2$ msec
(short avalanches).
\ Low water flow rates are analogous to slowly driving the system towards
the threshold of instability, as in other avalanche
studies\cite{sand,vortices}.  High flow rates drive the system very rapidly,
preventing a broad distribution of relaxation rates, resulting in exponential
distributions of sizes and lifetimes.

\pp
At low T and low flow rates, there were several somewhat periodic, large
events which were significantly larger than the largest events in the high T
data sets. These large events correspond to avalanches which span a
significant  portion of the dome
surface and sweep up many stationary droplets in their paths, thus acting
like ``system-resetting'' events.  After the occurrence of any one of these
large events, the behavior was characterized by only small and medium size
avalanches for a long period of time.  Eventually the system collected
enough stationary droplets to have the potential for another system-resetting
avalanche.  This behavior is reminiscent of the relaxation oscillations
in sandpiles\cite{sand}, where every ``period'' exhibits precursor
avalanches leading up to a large system-spanning slide, which resets and
relaxes
the system, followed by a series of small aftershocks.  The large water
avalanches occurred because the higher viscosity obtained at low
T provided more ``cohesion'' between the droplets, allowing
them to accrete into much larger clusters on the surface before avalanching.

\pp
The distributions, $D(\Delta t)$, of delay time intervals between successive
events
shown in figure 3 exhibit a stretched exponential behavior:
$D(\Delta t) \sim \exp\{-(\Delta t/\Delta t_0)^{1/4} \} $, where
$\Delta t_0 = 2.8$ msec (for the low T data shown).
\ The $D(\Delta t)$ for the low T, high viscosity data has a higher
$\Delta t$ tail
than the corresponding tail for the high T, low viscosity data.
\ At any given T, we found that increasing the flow rate caused the
range of the $\Delta t$ tails to be compressed.
\ This is consistent with mass conservation: the more quickly water is sprayed
into the system, the more closely spaced the resulting avalanches will be.
\ The small exponent in the fit of the high viscosity data suggests that
correlations among droplet avalanches are significant.

\pp
\ Figure 4 shows the power spectra obtained by squaring the Fourier
transform of the avalanche time trace.
\ These power spectra, with three different ``regions'', are similar to those
published in some previous theoretical avalanche studies\cite{hwakardar}.  At
higher viscosities (low T),
one might identify three approximate power-law regimes:
a high-frequency regime ($f \geq 90$ Hz) with a slope of about $-2.7$,
corresponding to observation times smaller than our
largest lifetime avalanche;
an intermediate regime ($30 \leq f < 90$ Hz) with a slope of the order of $-1$,
corresponding to times larger than the maximum duration (lifetime)
of individual avalanches ($\sim 13$ msec), and denoted\cite{hwakardar}
as the regime of interacting avalanches; and a low-frequency regime
($f < 30$ Hz, for low T) with an approximately flat slope, characteristic
of white noise, and corresponding to uncorrelated  events widely separated
in time.  For the low viscosity data (high T), the middle-frequency regime
is small, while the (low-frequency) white noise region extends significantly
(up to $f \sim 40$ Hz).
\ This demonstrates that for lower viscosities, the avalanches are more
weakly correlated than the higher viscosity events.  Of course,
and as discussed by Weissman, Jensen, and others\cite{weissman}, any
occurrence of a $1/f$ region by itself should not be
considered definite evidence in support of self-organized criticality.

\pp
Our distributions demonstrate that avalanches with many different sizes and
time delays occur.  To examine if there are significant correlations
between consecutive avalanches, we have studied the $\Delta t$ %and $S$
return maps.
\ Fig.~5 shows $\Delta t_{n+1}$ versus $\Delta t_{n}$ for the high viscosity
(low T) data set for two flow rates: (a) $17$ cc/min, and (b) $8$ cc/min.
\ In previous dripping faucet chaos experiments\cite{faucets}, this graph
produced ``periodic'' islands with interesting fractal shapes and some
purely stochastic distributions, illustrating period-doubling behavior
as a function of the faucet flow rate.  However, in our return
map, there is a fundamentally different behavior.  Instead of periodic
islands, there is a high concentration along the axes and a dense
triangular region at the origin.
\ For high flow rates, event pairs with a large time delay are usually
followed by another avalanche a short time later.  Conversely, event pairs
with smaller time delays are followed by an avalanche with an increasingly
larger range of time delay possibilities.
\ However, for low flow rates, the return map is more uniformly distributed
due to the broader distribution of delay times.
\ The observed behavior is the result of the many degrees
of freedom involved in the dynamics, because this system can be thought of
as a large array of interacting dripping faucets.

\pp
Very large
avalanches (e.g., $D(S=17)$ in fig.2(a)) are rare, and thus are more
affected by statistical fluctuations than the smaller size events.
\ To improve our statistics (e.g., to minimize fluctuations)
and increase the range of values for $S$, $\tau$, and $\Delta t$
significantly, several alterations would be required:
a much larger sprayed surface, longer measurement times,
and lower flow rates.
\ However, experimental and computer limitations prevented us from
making these changes.

\pp
In summary,
the general motivation of this work is two-fold: the physics of continuously
driven water droplet avalanches has not been systematically explored so far,
and more specifically, the new paradigm of critical dynamical attractors,
studied in a variety of different systems, has not yet been examined
in detail for fluids.
\ At both high and low viscosities and at low flow rates ($\sim 8$ cc/min)
we have found avalanche scale invariance in analogy with other non-fluid
systems.  Furthermore, higher viscosities provided a larger range of power
law behavior due to the increased ``cohesive strength'' (i.e., viscosity)
between the droplets.
\ At high flow rates ($\sim 17$ cc/min), the event distributions did
not exhibit scale invariance, but rather decayed exponentially, implying
a subcritical state with characteristic
length and time scales.  By lowering the input current (i.e., flow rate),
the system moved from a subcritical regime to a state with no
characteristic scales.

\pp
This work was supported in part by NSF grant DMR-90-01502 (FN) and a
Research Opportunity Award from the Research Corporation (MB).  BP received
support for undergraduate students from the NSF-REU.

\newpage
\input droplet-ref.tex  %%%  <<---THIS IS A SEPARATE FILE WITH REFERENCES **
%%%%%%%%%%%%%%%%%%%%%%%%%%%       IT IS ATTACHED AT THE END ****************

\newpage

\topmargin -0.6in
\textheight 9.6in
\vspace*{-0.3in}
{\large \bf Figure captions}
\vskip -0.05in

Fig.~1  \ (a) Schematic diagram of the experimental apparatus:
distilled water is sprayed
through the spray mister (A) into the transparent plastic (plexiglass) dome
(B).  Streams then run down the dome, drop onto the slanted annular impact
ring (C), %, which is a section of a cone.
and immediately drop off the inner rim of the ring.
\ Their impulse onto the ring produces signals from the three piezo film
detectors (D), mounted at $120^{\circ}$ intervals,
one of which is not shown
for clarity.  (b) Detailed diagram of a piezo film detector: upon a droplet
impulse, the suspension wire (E) stretches the piezo film strip (F),
enclosed in a copper shield (G), sending an electric signal to the computer
through %BNC
coaxial cables (H).

\pp
Fig.~2 \ Distributions of water droplet avalanche sizes, $D(S)$, (a) and
lifetimes, $D(\tau)$, (b) for two 3.5 hour runs
with a flow rate of $8$ cc/min at two different viscosities (temperatures):
low viscosity ($\sim 22^{\circ}$C) ($\bigtriangleup$) producing 2569
avalanches; and high viscosity ($\sim1^{\circ}$C)
($\bullet$) producing 2961 avalanches.
In this paper, open triangles ($\bigtriangleup$) always correspond to high T
data, while filled circles ($\bullet$) represent low T data.
\ $S$ is dimensionless since it is obtained as a ratio of voltages.

\pp
Fig.~3 \
Distributions of delay times between successive avalanches, $D(\Delta t)$,
for the same data described in fig.~2

\pp
Fig.~4 \
Power spectrum versus frequency (in Hertz) for the same data shown in fig.~2.
\ Straight lines denote slopes of -1 and -2.7.  For clarity, data has not been
normalized.

\pp
Fig.~5 \
Return maps, $\Delta t_{n+1}$ versus $\Delta t_{n}$, for high viscosity
data (low T) at two different flow-rates:
(a) 17 cc/min, and (b) 8 cc/min.  A steep
decay in $D(\Delta t)$ for high flow-rates produces a dense cluster of data
around a triangular region near the origin.  A slowly decaying $D(\Delta t)$
for low flow-rates produces a much more diffuse distribution of data.

\end{document}